\documentclass[preprint,aps]{revtex4}
\usepackage{graphicx}
\usepackage{booktabs}
\usepackage[usenames, dvipsnames]{color}
\usepackage{multirow}
\usepackage{setspace}
\usepackage{amsmath, amsthm, amssymb}
\usepackage{rotating}
\usepackage[ pdftex, plainpages = false, pdfpagelabels, 
                 pdfpagelayout = useoutlines,
                 bookmarks,
                 bookmarksopen = true,
                 bookmarksnumbered = true,
                 breaklinks = true,
                 linktocpage=all,
                 pagebackref=false,
                 colorlinks = true,
                 linkcolor = BrickRed,
                 urlcolor  = blue,
                 citecolor = BrickRed,
                 anchorcolor = green,
                 hyperindex = true,
                 hyperfigures
                 ]{hyperref} 
                 
\newcommand{\comment}[1]{}

\begin{document}

\title{Food for Fuel: The Price of Ethanol}
\author{Dominic K. Albino, Karla Z. Bertrand, and \href{http://necsi.edu/faculty/bar-yam.html}{Yaneer Bar-Yam}}
\affiliation{\href{http://www.necsi.edu}{New England Complex Systems Institute} \\ 
238 Main St. Suite 319 Cambridge MA 02142, USA \vspace{2ex}}
\date{October 5, 2012}

\begin{abstract}
Conversion of corn to ethanol in the US since 2005 has been a major cause
of global food price increases during that time and has been shown
to be ineffective in achieving US energy independence and reducing
environmental impact. We make three key statements to enhance understanding
and communication about ethanol production\textquoteright{}s impact
on the food and fuel markets: (1) The amount of corn used to produce
the ethanol in a gallon of regular gas would feed a person for a day,
(2) The production of ethanol is so energy intensive that it uses only
20\% less fossil fuel than gasoline, and (3) The cost of gas made
with ethanol is actually higher per mile because ethanol reduces gasoline\textquoteright{}s
energy per gallon.
\end{abstract}

\maketitle
\section*{Introduction}

The US used over 45\% of its 2011 corn crop to produce ethanol, up
from under 15\% before 2005 \cite{USDAERScorn}--a rise dictated by
federal mandate and promoted by federal subsidies. The drought in
2012 is leading to questions about whether using corn for fuel is
reasonable while people go hungry due to a world food shortage. 
Here we provide a context for this discussion by addressing the 
following three questions: How much food goes into ethanol 
production? What is the net gain in energy from ethanol 
production after accounting for the fossil fuel energy used 
in the process? How does the use of ethanol affect gasoline
prices?

\section*{1. Food used to make ethanol}

How many people could be fed with the corn used for ethanol? In the
US, the primary ethanol production input is field corn. While not
typically eaten on the cob, field corn is used to make other food
including breakfast cereal, corn flour or meal, corn sweetener, and
corn oil. It serves both as food for people and feed for
livestock--which subsequently becomes food for people in the form of poultry, pork, and beef. 

A bushel of field corn can be used to produce about 2.77 gallons of
ethanol \cite{Donner2008,Korves2008,USDAMNS2012,FAPRI2006,Eckhoff2007}.
A bushel of field corn weighs 56 pounds, each pound containing about
1,550 Calories \cite{Dale1994,Cromwell2006}. Therefore, it takes
about 31,300 Calories of field corn to produce one gallon of ethanol.
Regular gasoline (E10) typically contains 10\% ethanol by volume (averaging
9.6\% nationally in 2011 \cite{EIAoutlook}). Therefore, about 3,000 Calories 
of corn energy is used to produce each gallon of regular gas. 

The suggested daily food energy intake is 2,100 Calories per person
\cite{WFP2012}. A single gallon of regular gas contains more than
enough food energy to feed a person for one day. More precisely, every
gallon would feed 1.4 people for a day or one person for 1.4 days.

It is often pointed out that a portion of the corn energy does not
end up in the ethanol, but instead in by-products subsequently fed
to animals, called distillers grains, which may account for up to
31\% of the corn by weight \cite{USDAERSbackground}. The actual amount
is often less than 31\% and in 2011 was 23\% overall \cite{WASDE2012}.
The by-products are not used for food, and their nutritional content
limits their use for animal feed \cite{DiCostano2005}. 
Even after removing 23\% or 31\% of its food energy, a gallon of
regular gas still contains more than enough to feed a person for a day.

What is the equivalent amount of sweet (``on the cob'') corn in regular gas? The energy in sweet
corn is 485 Calories per pound \cite{Self2012,CalorieLab2012}, and there are
about 0.2-0.25 edible pounds on each ear of corn \cite{Self2012}.
In a gallon of regular gas there is food energy equivalent to 28 ears 
of sweet corn, while just 19 ears of sweet corn would satisfy a person's
daily energy requirement. 

An average 16 gallon tank of gas contains ethanol from enough corn to feed 22 
people for a day, or one person for over three weeks. 

The total amount of ethanol produced in the US in 2011 was 13.95 billion gallons \cite{EIAoutlook},
enough to feed 570 million people that year.

\section*{2. Energy Balance of Corn Ethanol}

When weighing the costs and benefits of corn ethanol, it is important
to consider the net energy yield: how much more energy we get from
a gallon of ethanol than is used to make that gallon. Fossil fuel energy is required
to produce and transport fertilizers and pesticides, irrigate farmland,
and plant and harvest the corn (not including the solar energy involved).
Additional energy is required to transport the corn from the field
to the ethanol plants and power the conversion process. 

The most optimistic assessments claim around $1.3$ units of energy are 
produced for each unit of energy input \cite{DOEmyths}.
However, this estimate uses data from the best corn-growing conditions
(requiring
relatively low costs to grow), and considers the best processing conditions
(including the highest possible distiller grain outputs as well).
The average of the net energy yield across all corn-growing regions
has been found to be $1.01$ \cite{Murphy2010}.
This means that the same amount of fossil fuel energy goes into making
a gallon of ethanol as can be obtained by using that ethanol in a car. 

Gasoline also requires energy to make. However, the ratio of output
to processing energy for petroleum is about 5 to 1 \cite{DOEmyths}. This
means that an extra $20\%$ fossil fuel is used for each gallon of gasoline
that is burned in a car \cite{Rapier2006,Hofstrand2009,DOEargonne}. 

Thus we need $1.0$ fossil fuel BTUs on average to produce $1.0$ ethanol
BTUs of fuel energy. 
We need $1.2$ fossil fuel BTUs to produce $1.0$ fossil fuel BTUs of
fuel energy. By choosing ethanol instead of gasoline, we save about
$0.2$ fossil fuel BTUs for each BTU used. This is the only gain from 
ethanol after loss of its food value. The amount of fossil fuel saved 
by using ethanol is only about $0.2\%$ of the US energy requirements 
\cite{energy_calc1,energy_calc2}.

\section*{3. Gasoline Price Benefit of Corn Ethanol}

While it has been claimed that ethanol has reduced the price of gasoline \cite{Du2011,Du2012},
what is reported is the cost per gallon, but what is relevant is the cost per mile driven. Ethanol
has less energy per gallon than gasoline. A gallon of gasoline 
contains about 125,000 BTUs while ethanol contains about 84,300 BTUs \cite{EIAenergycontent}, or
about 67\% that of gasoline. When the price of ethanol is between 67\% and 100\% of the price of
gasoline, which it often is, ethanol is cheaper by volume but more expensive by energy. 
The cost per gallon of gasoline with ethanol is lower, but it is as if the gasoline is watered down---the 
cost per mile driven is higher. 

This means that in addition to the government subsidy of \$20 billion from 2005-2011 \cite{NPR2012}, every gallon
of gasoline with ethanol bought is an extra subsidy from consumers to the ethanol producers.


\begin{thebibliography}{10}

\bibitem{USDAERScorn}
US domestic corn use, {\it USDA Economic Research Service\/}  (2012
  \url{http://www.ers.usda.gov/media/866543/cornusetable.html}).

\bibitem{Donner2008}
S.~D. Donner, C.~J. Kucharik, Corn-based ethanol production compromises goal of
  reducing nitrogen export by the Mississippi River, {\it PNAS\/} {\bf 105},
  4513 (2008 \url{http://www.pnas.org/content/105/11/4513.full.pdf}).

\bibitem{Korves2008}
R.~Korves, The potential role for corn ethanol in meeting the energy needs of
  the United States in 2016-2030, {\it Illinois Corn Marketing Board\/}  (2008
  \url{http://www.ilcorn.org/uploads/useruploads/files/ethanol/potential_role_for_corn_ethanol.pdf}).

\bibitem{USDAMNS2012}
J.~Inman, Iowa ethanol corn and co-products processing values, {\it USDA Market
  News Service\/}  (2012 \url{http://www.ams.usda.gov/mnreports/nw_gr212.txt}).

\bibitem{FAPRI2006}
Biofuel conversion factors, {\it Food and Agricultural Policy Research
  Institute, University of Missouri\/}  (2006
  \url{http://www.fapri.missouri.edu/outreach/publications/2006/biofuelconversions.pdf}).

\bibitem{Eckhoff2007}
S.~R. Eckhoff, Choosing a fractionation process, {\it Maize Processing
  Innovators, Inc.\/}  (2007
  \url{http://www.card.iastate.edu/publications/dbs/pdffiles/11wp523.pdf}).

\bibitem{Dale1994}
N.~Dale, Relationship between bushel weight, metabolizable energy, and protein
  content of corn from an adverse growing season, {\it Journal of Applied
  Poultry Research\/} {\bf 3}, 83 (1994
  \url{http://japr.fass.org/content/3/1/83.full.pdf}).

\bibitem{Cromwell2006}
G.~L. Cromwell, Benefits of high oil corn for swine, {\it University of
  Kentucky College of Agriculture\/}  (2006
  \url{http://www.uky.edu/Ag/AnimalSciences/pubs/highoilcornbenefitsforswine.pdf}).

\bibitem{EIAoutlook}
Short term energy outlook, {\it U.S. Energy Information Administration\/}
  (2012 \url{http://www.eia.gov/forecasts/steo/archives/aug12.pdf}).

\bibitem{WFP2012}
What is hunger?, {\it World Food Programme\/}  (2012
  \url{http://www.wfp.org/hunger/what-is}).

\bibitem{USDAERSbackground}
Corn background, {\it USDA Economic Research Service\/}  (2012
  \url{http://www.ers.usda.gov/topics/crops/corn/background.aspx}).

\bibitem{WASDE2012}
{World Agricultural Outlook Board}, World agricultural supply and demand
  estimates, {\it USDA\/}  (2012
  \url{http://www.usda.gov/oce/commodity/wasde/latest.pdf}).

\bibitem{DiCostano2005}
A.~DiCostanzo, Feeding distillers grains to beef cattle, {\it University of
  Minnesota Extension Beef Center\/}  (2005
  \url{http://www.extension.umn.edu/beef/components/releases/02-28-05-DiCostanzo.htm}).

\bibitem{Self2012}
Nutrition Facts: Corn, sweet, yellow, cooked, boiled, drained, without salt,
  {\it Self Nutrition Data\/}  (2012
  \url{http://nutritiondata.self.com/facts/vegetables-and-vegetable-products/2416/0}).

\bibitem{CalorieLab2012}
Nutrition Facts: Corn, sweet, yellow, cooked, boiled, drained, without salt,
  {\it CalorieLab.com\/}  (2012
  \url{http://calorielab.com/foods/corn/corn-sweet-yellow-cooked-boiled-drained-without-salt/148/11168/4}).

\bibitem{DOEmyths}
Ethanol myths and facts, {\it U.S. Department of Energy, Energy Efficiency and
  Renewable Energy Biomass Program\/}  (2011
  \url{http://www1.eere.energy.gov/biomass/printable_versions/ethanol_myths_facts.html}).

\bibitem{Murphy2010}
D.~J. Murphy, C.~A.~S. Hall, B.~Powers, New perspectives on the energy return
  on (energy) investment (EROI) of corn ethanol, {\it Environment, Development,
  and Sustainability\/} {\bf 13}, 179 (2010
  \url{http://www.springerlink.com/content/j458318434015735/fulltext.pdf}).

\bibitem{Rapier2006}
R.~Rapier, The energy balance of ethanol versus gasoline, {\it The Oil Drum\/}
  (2006 \url{http://www.theoildrum.com/story/2006/8/25/221617/881}).

\bibitem{Hofstrand2009}
D.~Hofstrand, Efficiency and environmental improvements of corn ethanol
  production, {\it Agricultural Marketing Resource Center Renewable Energy
  Newsletter\/}  (2009
  \url{http://www.agmrc.org/renewable_energy/ethanol/efficiency-and-environmental-improvements-of-corn-ethanol-production/}).

\bibitem{DOEargonne}
Argonne National Laboratory ethanol study: Key points, {\it U.S. Department of
  Energy, Office of Renewable Energy and Energy Efficiency\/}  (2005
  \url{http://www.ethanolmt.org/images/argonnestudy.pdf}).

\bibitem{energy_calc1}
Annual energy review table 10.3: Fuel ethanol overview, 1981-2010, {\it US
  Energy Information Administration\/}  (2011
  \url{http://www.eia.gov/totalenergy/data/annual/showtext.cfm?t=ptb1003}).

\bibitem{energy_calc2}
Energy overview: Total energy flow, 2010, {\it US Energy Information
  Administration\/}  (2010
  \url{http://www.eia.gov/totalenergy/data/annual/diagram1.cfm}).

\bibitem{Du2011}
X.~Du, D.~J. Hayes, The impact of ethanol production on US and regional
  gasoline markets: An update to May 2009, {\it Center for Agricultural and
  Rural Development, Iowa State University\/} {\bf 11-WP 523} (2011
  \url{http://www.card.iastate.edu/publications/dbs/pdffiles/11wp523.pdf}).

\bibitem{Du2012}
X.~Du, D.~J. Hayes, The impact of ethanol production on U.S. and regional
  gasoline markets: An update to 2012, {\it Center for Agricultural and Rural
  Development, Iowa State University\/} {\bf 12-WP 528} (2012
  \url{http://www.card.iastate.edu/publications/dbs/pdffiles/12wp528.pdf}).

\bibitem{EIAenergycontent}
Biofuels in the U.S. transportation sector, {\it U.S. Energy Information
  Administration\/}  (2007
  \url{http://www.eia.gov/oiaf/analysispaper/biomass.html}).

\bibitem{NPR2012}
Congress ends era of ethanol subsidies, {\it National Public Radio (NPR)
  Morning Edition\/}  (January 3, 2012
  \url{http://www.npr.org/2012/01/03/144605485/congress-ends-era-of-ethanol-subsidiesm}).

\end{thebibliography}
\end{document}